\documentclass[aip,graphicx,secnumroman,eqsecnum]{revtex4-2}
\usepackage{graphicx}
\usepackage{dcolumn}
\usepackage{bm}

\newcommand{\beq}{\begin{equation}}
\newcommand{\eeq}{\end{equation}}
\newcommand{\beqa}{\begin{eqnarray}}
\newcommand{\eeqa}{\end{eqnarray}}
\newcommand{\vc}[1]{\mbox{\boldmath $#1$}}

\newcommand{\vol}[1]{{\bf #1}}

\newcommand{\du}[1]{{\bf\sf #1}}

\begin{document}


\title{Transition in steady streaming and pumping caused by a sphere oscillating in a viscous incompressible fluid}

\author{B. U. Felderhof}
\thanks{In memory of Rudi Schmitz, a great friend}
\email{ufelder@physik.rwth-aachen.de}
\affiliation{Institut f\"ur Theorie der Statistischen Physik \\ RWTH Aachen University\\
Templergraben 55\\52056 Aachen\\ Germany\\}

\date{\today}

\begin{abstract}
The steady streaming flow pattern caused by a no-slip sphere oscillating in an unbounded viscous incompressible fluid is calculated exactly to second order in the amplitude. The pattern depends on a dimensionless scale number, determined by sphere radius, frequency of oscillation, and kinematic viscosity of the fluid. At a particular value of the scale number there is a transition with a reversal of flow. The analytical solution of the flow equations is based on a set of antenna theorems. The flow pattern consists of a boundary layer and an adjacent far-field of long range, falling off with the inverse square distance from the center of the sphere. The boundary layer becomes
thin in the limit where inertia dominates over viscosity. The system acts as a pump operating in two directions, depending on the scale number. The efficiency of the pump is estimated from a comparison of the rate of flow with the rate of dissipation.
\end{abstract}

\maketitle\
\section{\label{I}Introduction}

The phenomenon of steady streaming is of fundamental interest in fluid mechanics \cite{1,2,3}. It is a manifestation of the Reynolds stress in a moving fluid corresponding to the convection of momentum, as described by the nonlinear convective term in the Navier-Stokes equations.
Oscillating boundaries cause waves whose interference leads to steady effects by rectification. Well-known examples are Kundt's experiment, as explained by Rayleigh\cite{4}, and the formation of sandbars, as explained by Bagnold and Taylor \cite{5}. A photograph of the steady flow pattern caused by a vibrating cylinder is presented in the Album of Fluid Motion \cite{6}.

The simple geometry of a sphere with no-slip boundary condition oscillating in unbounded fluid is of prime interest. In a series expansion of the steady streaming velocity in powers of the amplitude of oscillation the first non-vanishing term is of second order. We restrict attention to this lowest order term.

The relative importance of viscous and inertial effects is measured by a dimensionless scale number $s=a\sqrt{\omega\rho/2\eta}$, where $a$ is the radius of the sphere, $\omega$ is the frequency, $\rho$ is the mass density of the fluid, and $\eta$ is its shear viscosity. The product $4\varepsilon s^2$ is sometimes called the Reynolds number Re, where $\varepsilon$ is the ratio of amplitude $A$ and sphere radius $a$, and $\mathrm{Re}_s=\varepsilon\mathrm{Re}$, is called the streaming Reynolds number.

Wang \cite{7} derived an approximate expression for the steady streaming velocity based on matched asymptotic expansions. Riley \cite{8}  derived expressions valid in the limits of small and large scale number $s$. Amin and Riley\cite{9} considered a slightly different, but related, geometry, and derived numerical results in the intermediate regime. They use the variable $M=2s$. Riley \cite{10} reviewed the earlier work.

Voth et al. \cite{11} showed how the single sphere result can be used to find a hydrodynamic pair interaction in a vibrated colloidal suspension. Klotsa and Swift \cite{12} studied both the single sphere flow pattern and the pair interaction in more detail by computer simulation. Otto et al.\cite{13} performed measurements on the single sphere flow pattern, but found comparison with then existing theory difficult. Spelman and Lauga \cite{14} rederived Riley's result for large scale number by a different method and generalized it to other boundary conditions. They discuss a variety of physical situations in which the calculation is useful. Recently, van Overveld et al. \cite{15} studied the pair interaction by numerical simulation.

In the following we find the analytic solution of the single sphere problem to second order in the amplitude  of oscillation for no-slip boundary conditions. We use the method derived by Felderhof and Jones\cite{16} for the swimming problem to show that one requires the solution of a boundary value problem for the solution of the steady-state homogeneous Stokes equations, as well as a solution of the nonhomogeneous Stokes equations with the Reynolds stress as source. It is natural to solve the latter problem by the method of Green functions. In earlier work \cite{17,18} we derived a set of antenna theorems valid for unbounded fluid. Here we employ the generalization to infinite fluid with a fixed immersed sphere with no-slip boundary condition \cite{19}.

It turns out that the antenna theorems provide a useful tool, but the explicit analytic form of the mean flow velocity, $\overline{\vc{u}}(r,\theta)$ in spherical coordinates in terms of known functions, is quite unwieldy. However, the derived integral form leads to fast and accurate numerical results. In particular we recover Riley's result at high frequency. At low frequency he used an asymptotic matching method which is not required for our solution.

We study in particular the flow velocity along the vertical axis $\overline{\vc{u}}(r,0)$ and along the horizontal axis $\overline{\vc{u}}(r,\pi/2)$. The behavior for various values of the scale number $s$ makes clear that flow reversal occurs at $s_0=2.85632$. The corresponding critical Reynolds number is $\mathrm{Re}_c=4\varepsilon s_0^2=16.1346\varepsilon$. If the system is regarded as a pump, then for $s<s_0$ it pumps fluid from the polar directions $\theta=0$ and $\theta=\pi$ into the equatorial plane $\theta=\pi/2$. For $s>s_0$ the flow  is reversed.

\section
{\label{II}Steady streaming about a sphere}

We consider a sphere of radius $a$  immersed in a viscous incompressible fluid of shear viscosity $\eta$ and mass density $\rho$. The fluid is of infinite extent in all directions and the sphere performs small harmonic oscillations about a point $O$, which is taken to be the origin of a Cartesian system of coordinates. The oscillations are along a line which is taken to be the $z$-axis. Thus the position of the sphere center is given by
 \begin{equation}
\label{2.1}\vc{\xi}(t)=\varepsilon a\sin(\omega t)\vc{e}_z,
\end{equation}
where $\varepsilon a$ is the amplitude of oscillations, $\omega$ is the frequency, and $\vc{e}_z$ is the unit vector in the $z$-direction. The linear velocity of the sphere is
\begin{equation}
\label{2.2}\vc{U}(t)=\varepsilon \omega a\cos(\omega t)\vc{e}_z.
\end{equation}
We assume that the fluid velocity satisfies the no-slip boundary condition at the surface of the sphere. The flow velocity $\vc{u}(\vc{r},t)$ and pressure $p(\vc{r},t)$ are assumed to satisfy the Navier-Stokes equations
 \begin{equation}
\label{2.3}\rho\bigg{[}\frac{\partial\vc{u}}{\partial t}+\vc{u}\cdot\nabla\vc{u}\bigg{]}=\eta\nabla^2\vc{u}-\nabla p,\qquad\nabla\cdot\vc{u}=0.
\end{equation}
We are interested in the mean flow velocity $\overline{\vc{u}}(\vc{r})$ and the mean pressure $\overline{p}(\vc{r})$, defined as the averages over a period $T=2\pi/\omega$ in the stationary periodic state of the system. These averages will be calculated exactly to second order in $\varepsilon$ by the method of Felderhof and Jones\cite{16}, with the change of frame as discussed in Ref. 19.

The mean second order fields $\overline{\vc{u}^{(2)}}(\vc{r})$ and $\overline{p^{(2)}}(\vc{r})$ satisfy the steady state Stokes equations
 \begin{equation}
\label{2.4}\eta\nabla^2\overline{\vc{u}^{(2)}}-\nabla\overline{p^{(2)}}=-\overline{\vc{F}_R^{(2)}} ,\qquad\nabla\cdot\overline{\vc{u}^{(2)}}=0,
\end{equation}
with mean Reynolds force density
\begin{equation}
\label{2.5}\overline{\vc{F}_R^{(2)}}=-\rho\overline{\vc{u}^{(1)}\cdot\nabla\vc{u}^{(1)}},
\end{equation}
and boundary condition
\begin{equation}
\label{2.6}\overline{\vc{u}^{(2)}}\big|_{r=a}=-\overline{\vc{\xi}\cdot\nabla\vc{u}^{(1)}}\big|_{r=a}.
\end{equation}
The mean fields can be written as a sum of two terms
\begin{equation}
\label{2.7}\overline{\vc{u}^{(2)}}=\overline{\vc{u}^{(2)}}_S+\overline{\vc{u}^{(2)}}_B,\qquad
\overline{p^{(2)}}=\overline{p^{(2)}}_S+\overline{p^{(2)}}_B,
\end{equation}
with surface term $\overline{\vc{u}^{(2)}}_S$ satisfying Eq. (2.4) with vanishing right-hand side and boundary condition Eq. (2.6), and with bulk term $\overline{\vc{u}^{(2)}}_B$ satisfying Eq. (2.4) with no-slip boundary condition. First we discuss the surface contribution.

\section{\label{III}Surface contribution}

The surface contribution $\overline{\vc{u}^{(2)}}_S(\vc{r})$ is easily evaluated. In complex notation the first order fields take the form
\begin{equation}
\label{3.1}\vc{u}^{(1)}(\vc{r},t)=\mathrm{Re}[\vc{u}_\omega(\vc{r})e^{-i\omega t}],\qquad p^{(1)}(\vc{r},t)=\mathrm{Re}[p_\omega(\vc{r})e^{-i\omega t}],
\end{equation}
with amplitudes $\vc{u}_\omega(\vc{r}),p_\omega(\vc{r})$
which satisfy the linearized Navier-Stokes equations
\begin{equation}
\label{3.2}\eta[\nabla^2\vc{u}_\omega-\alpha^2\vc{u}_\omega]-\nabla p_\omega=0,\qquad\nabla\cdot\vc{u}_\omega=0,
\end{equation}
with the variable
\begin{equation}
\label{3.3}\alpha=(-i\omega\rho/\eta)^{1/2}=(1-i)(\omega\rho/2\eta)^{1/2}.
\end{equation}
The relevant first order solution with no-slip boundary condition is \cite{21,22}
\begin{equation}
\label{3.4}\vc{u}^{(1)}_{\omega}(\vc{r})=\frac{1}{2}\varepsilon \omega a\big[\alpha a\vc{v}_1(\vc{r},\alpha)
+\frac{k_2(\alpha a)}{k_0(\alpha a)}\vc{u}_1(\vc{r})\big],
\end{equation}
with modified spherical Bessel functions $k_l(z)$ and modes
 \begin{eqnarray}
\label{3.5}\vc{v}_1(\vc{r},\alpha)&=&\frac{2}{\pi}\;e^{\alpha a}[2k_0(\alpha r)\vc{e}_z+k_2(\alpha r)(\vc{e}_z-3\cos\theta\;\vc{e}_r)],\nonumber\\
\vc{u}_1(\vc{r})&=&-\bigg(\frac{a}{r}\bigg)^3(\vc{e}_z-3\cos\theta\;\vc{e}_r),\qquad p_1(\vc{r},\alpha)=\eta\alpha^2a\bigg(\frac{a}{r}\bigg)^2\cos\theta,
\end{eqnarray}
with radial unit vector $\hat{\vc{r}}=\vc{r}/r=\vc{e}_r$ and polar angle $\theta$. Hence the mean surface velocity, given by Eq. (2.6), is evaluated as
\begin{equation}
\label{3.6}\overline{\vc{u}^{(2)}}_S\big|_{r=a}=\frac{3}{8}\varepsilon^2\;s\;\omega\;a\;\sin(2\theta)\;\vc{e}_\theta,
\end{equation}
where $s$ is the dimensionless scale number $s=a\sqrt{\omega\rho/(2\eta)}$.

In order to find the corresponding solution of the Stokes equations we recall the set of axisymmetric solutions defined in Ref. 19 as
\begin{eqnarray}
\label{3.7}
\vc{v}^0_l(\vc{r})&=&\bigg(\frac{a}{r}\bigg)^l\bigg[\frac{2l+2}{l(2l+1)}\vc{A}_l-\frac{2l-1}{2l+1}\vc{B}_l\bigg],\nonumber\\
\vc{u}_l(\vc{r})&=&-\bigg(\frac{a}{r}\bigg)^{l+2}\vc{B}_l(\hat{\vc{r}}),
\end{eqnarray}
with vector spherical harmonics $\{\vc{A}_l,\vc{B}_l\}$ defined by Cichocki et al.\cite{23}
 \begin{eqnarray}
\label{3.8}\vc{A}_l&=&\hat{\vc{A}}_{l0}=lP_l(\cos\theta)\vc{e}_r-P^1_l(\cos\theta)\vc{e}_\theta,\nonumber\\
\vc{B}_l&=&\hat{\vc{B}}_{l0}=-(l+1)P_l(\cos\theta)\vc{e}_r-P^1_l(\cos\theta)\vc{e}_\theta,
\end{eqnarray}
with Legendre polynomials $P_l$ and associated Legendre functions $P^1_l$ in the notation of Edmonds\cite{24}.

Fitting the boundary condition Eq. (3.6) we find that the required solution of the Stokes equations is
\begin{equation}
\label{3.9}\overline{\vc{u}^{(2)}_S}(\vc{r})=
\frac{-1}{4}\;\varepsilon^2\;s\;a\;\omega\;[\vc{v}^0_2(\vc{r})-\vc{u}_2(\vc{r})].
\end{equation}
The expression implies that the surface contribution to the flow is the sum of an electrostatic quadrupole field and a radial field with angular dependence given by $P_2(\cos\theta)$. In the notation of Ref. 19 the flow is characterized by the two moments
\begin{equation}
\label{3.10}\kappa_{2S}=\;\frac{1}{4}\varepsilon^2\;s,\qquad\mu_{2S}=-\frac{1}{4}\varepsilon^2\;s.
\end{equation}
Both moments diverge in the limit of large $s$. We show in the following that this behavior is canceled by a corresponding contribution from the Reynolds force density in the bulk.

\section{\label{IV}Antenna theorems}

In order to calculate the  bulk contribution we must solve the inhomogeneous Stokes equations (2.4). To that purpose we derive antenna theorems involving integrals over a spherical surface $S(b)$ of radius $b$ centered at the origin, surrounding the sphere of radius $a$. Earlier we derived such theorems for a uniform fluid in the absence of the sphere\cite{17},\cite{18}. For that case the relevant identities read

 \begin{eqnarray}
\label{4.1}\int_{S(b)}\du{T}_0(\vc{r}-\vc{r}')\cdot\vc{A}_l(\hat{\vc{r}'})dS'
&=&4\pi\frac{l+1}{(2l-1)(2l+1)}\frac{r^{l-1}}{b^{l-2}}\vc{A}_l(\hat{\vc{r}}),\qquad{r<b},\nonumber\\
&=&4\pi\frac{l+1}{(2l-1)(2l+1)}\frac{b^{l+1}}{r^l}\vc{A}_l(\hat{\vc{r}})\nonumber\\
&+&2\pi\frac{l}{2l+1}\bigg(\frac{b^{l+3}}{r^{l+2}}-\frac{b^{l+1}}{r^l}\bigg)\vc{B}_l(\hat{\vc{r}}),\qquad r>b,\nonumber\\
\int_{S(b)}\du{T}_0(\vc{r}-\vc{r}')\cdot\vc{B}_l(\hat{\vc{r}'})dS'
&=&4\pi\frac{l}{(2l+1)(2l+3)}\frac{r^{l+1}}{b^l}\vc{B}_l(\hat{\vc{r}})\nonumber\\
&+&2\pi\frac{l+1}{2l+1}\bigg(\frac{r^{l+1}}{b^l}-\frac{r^{l-1}}{b^{l-2}}\bigg)\vc{A}_l(\hat{\vc{r}}),\qquad{r<b},\nonumber\\
&=&4\pi\frac{l}{(2l+1)(2l+3)}\frac{b^{l+3}}{r^{l+2}}\vc{B}_l(\hat{\vc{r}}),\qquad r>b,
\end{eqnarray}
where $\du{T}_0(\vc{r})$ is the Oseen tensor given by
\begin{equation}
\label{4.2}\du{T}_0(\vc{r})=\frac{1}{2r}(\vc{1}+\hat{\vc{r}}\hat{\vc{r}}).
\end{equation}
We write the expressions (4.1) in the form
 \begin{eqnarray}
\label{4.3}\frac{1}{4\pi}\int_{S(b)}\du{T}_0(\vc{r}-\vc{r}')\cdot\vc{A}_l(\hat{\vc{r}'})dS'
&=&[G_{0AAl<}(r,b)\Theta(b-r)+G_{0AAl>}(r,b)\Theta(r-b)]\vc{A}_l(\hat{\vc{r}})\nonumber\\
&+&G_{0BAl>}(r,b)\Theta(r-b)\vc{B}_l(\hat{\vc{r}}),\nonumber\\
\frac{1}{4\pi}\int_{S(b)}\du{T}_0(\vc{r}-\vc{r}')\cdot\vc{B}_l(\hat{\vc{r}'})dS'
&=&G_{0BBl<}(r,b)\Theta(b-r)\vc{B}_l(\hat{\vc{r}})\nonumber\\
&+&G_{0ABl<}(r,b)\Theta(b-r)]\vc{A}_l(\hat{\vc{r}})
+G_{0BBl>}(r,b)\Theta(r-b)\vc{B}_l(\hat{\vc{r}}),\nonumber\\
\end{eqnarray}
where $\Theta(x)$ is the Heaviside step-function. We modify this into corresponding theorems for the Green function of the Stokes equations for fluid outside a sphere of radius $a$ with flow satisfying the no-slip boundary condition.
The theorems take the form
 \begin{eqnarray}
\label{4.4}\int_{S(b)}\du{G}_a(\vc{r},\vc{r}')\cdot\vc{A}_l(\hat{\vc{r}'})dS'
&=&[G_{AAl<}(r,b)\Theta(b-r)+G_{AAl>}(r,b)\Theta(r-b)]\vc{A}_l(\hat{\vc{r}})\nonumber\\
&+&[G_{BAl<}(r,b)\Theta(b-r)+G_{BAl>}(r,b)\Theta(r-b)]\vc{B}_l(\hat{\vc{r}}),\nonumber\\
\int_{S(b)}\du{G}_a(\vc{r},\vc{r}')\cdot\vc{B}_l(\hat{\vc{r}'})dS'
&=&[G_{ABl<}(r,b)\Theta(b-r)+G_{ABl>}(r,b)\Theta(r-b)]\vc{A}_l(\hat{\vc{r}})\nonumber\\
&+&[G_{BBl<}(r,b)\Theta(b-r)+G_{BBl>}(r,b)\Theta(r-b)]\vc{B}_l(\hat{\vc{r}}).\nonumber\\
\end{eqnarray}
The tensor Green function on the left hand side tends to Oseen's form in the limit $a\rightarrow 0$,
\begin{equation}
\label{4.5}\lim_{a\rightarrow 0}\du{G}_a(\vc{r},\vc{r}') =\frac{1}{4\pi}\du{T}_0(\vc{r}-\vc{r}').
\end{equation}
The tensor Green function for arbitrary $a$ is known in explicit form\cite{25}, but this is not needed for the derivation of the following results.
We write the two vector functions on the right hand side of Eq. (4.4) as
 \begin{eqnarray}
\label{4.6}\vc{G}_{Al}(\vc{r})&=&\vc{G}_{0Al}(\vc{r})
+C_{Alu}\vc{u}_l(\vc{r})+C_{Alv}\vc{v}^0_l(\vc{r}),\nonumber\\
\vc{G}_{Bl}(\vc{r})&=&\vc{G}_{0lk Bl}(\vc{r})
+C_{Blu}\vc{u}_l(\vc{r})+C_{Blv}\vc{v}^0_l(\vc{r}).
\end{eqnarray}
The expressions guarantee that the functions have the correct behavior at infinity, and have the same jump in derivative at $r=b$ as the functions for uniform fluid. The coefficients $C$ can be determined from the no-slip
condition at $r=a$. Thus we find the radial functions \cite{19}
\begin{eqnarray}
\label{4.7}G_{AAl<}(r,b)&=&\frac{l+1}{4l^2-1}\frac{b^2}{b^lr^l}\big(r^{2l-1}-a^{2l-1}\big),\nonumber\\
 G_{AAl>}(r,b)&=&\frac{l+1}{4l^2-1}\frac{b^2}{b^lr^l}\big(b^{2l-1}-a^{2l-1}\big),\nonumber\\
G_{ABl<}(r,b)&=&\frac{l+1}{4l+2}\frac{1}{b^lr^{l}}\big(a^{2l-1}b^2-a^{2l+1}+r^{2l+1}-b^2r^{2l-1}\big),\nonumber\\
 G_{ABl>}(r,b)&=&\frac{l+1}{4l+2}\frac{a^{2l-1}}{b^lr^{l}}\big(b^2-a^2\big),\nonumber\\
G_{BAl<}(r,b)&=&\frac{l}{4l+2}\frac{a^{2l-1}b^2}{b^lr^{l+2}}\big(r^2-a^2\big),\nonumber\\
G_{BAl>}(r,b)&=&\frac{l}{4l+2}\frac{b^2}{b^lr^{l+2}}\big(a^{2l-1}r^2-a^{2l+1}+b^{2l+1}-b^{2l-1}r^2\big),\nonumber\\
G_{BBl<}(r,b)&=&\frac{l}{4(4l^2+8l+3)}\frac{1}{b^lr^{l+2}}
\big[4r^{2l+3}-(2l+1)^2a^{2l+3}+(4l^2+4l-3)\big(a^{2l+1}(b^2+r^2)-a^{2l-1}b^2r^2\big)\big],\nonumber\\
G_{BBl>}(r,b)&=&\frac{l}{4(4l^2+8l+3)}\frac{1}{b^lr^{l+2}}
\big[4b^{2l+3}-(2l+1)^2a^{2l+3}+(4l^2+4l-3)\big(a^{2l+1}(b^2+r^2)-a^{2l-1}b^2r^2\big)\big].\nonumber\\
\end{eqnarray}
In particular for $l=2$
\begin{eqnarray}
\label{4.8}G_{AA2<}(r,b)&=&\frac{r^3-a^3}{5r^2},\qquad G_{AA2>}(r,b)=\frac{b^3-a^3}{5r^2},\nonumber\\
G_{AB2<}(r,b)&=&\frac{3}{10b^2r^2}\big(r^5-a^5+a^3b^2-b^2r^3\big),\qquad G_{AB2>}(r,b)=\frac{3a^3}{10b^2r^2}(b^2-a^2),\nonumber\\
G_{BA2<}(r,b)&=&\frac{a^3(r^2-a^2)}{5r^4},\qquad G_{BA2>}(r,b)=\frac{b^5-a^5+b^3r^2-a^3r^2}{5r^4},\nonumber\\
G_{BB2<}(r,b)&=&\frac{4r^7-25a^7+21(a^5r^2+a^5b^2-a^3b^2r^2)}{70b^2r^4},\nonumber\\
G_{BB2>}(r,b)&=&\frac{4b^7-25a^7+21(a^5r^2+a^5b^2-a^3b^2r^2)}{70b^2r^4}.
\end{eqnarray}
Similar theorems can be derived for other boundary conditions and for bubbles or droplets. Analogous theorems were also derived in electrodynamics\cite{26} and in the theory of elasticity\cite{27}.

\section{\label{V}Bulk contribution}

In the calculation of the second order bulk contribution to the steady streaming we must solve the Stokes equations (2.4) with the time-averaged Reynolds force density $\overline{\vc{F}^{(2)}_R}(\vc{r})$, given by Eq. (2.5) , acting as a driving term. For the harmonic oscillation, given by Eq. (2.1), the first order flow velocity is a sum of terms,
\begin{equation}
\label{5.1}\vc{u}^{(1)}(\vc{r},t)=\vc{u}^{(1)}_c(\vc{r})\cos(\omega t)+\vc{u}^{(1)}_s(\vc{r})\sin(\omega t).                  \end{equation}
In complex notation $\vc{u}^{(1)}(\vc{r},t)=\mathrm{Re}[\vc{u}^{(1)}_\omega (\vc{r})\exp(-i\omega t)]$ with
\begin{equation}
\label{5.2}\vc{u}^{(1)}_c(\vc{r})=\mathrm{Re}[\vc{u}^{(1)}_\omega(\vc{r}) ],\qquad
\vc{u}^{(1)}_s(\vc{r})=\mathrm{Im}[\vc{u}^{(1)}_\omega(\vc{r}) ].
\end{equation}
The time-averaged Reynolds force density is given by
\begin{equation}
\label{5.3}\overline{\vc{F}^{(2)}_R}(\vc{r})=-\frac{1}{2}\rho\;\big[\vc{u}^{(1)}_c(\vc{r})\cdot\nabla\vc{u}^{(1)}_c(\vc{r})+
\vc{u}^{(1)}_s(\vc{r})\cdot\nabla\vc{u}^{(1)}_s(\vc{r})\big].
\end{equation}
The bulk contribution to the flow velocity is
\begin{equation}
\label{5.4}\overline{\vc{u}^{(2)}}_B(\vc{r})
=\frac{1}{\eta}\int_{r'>a}\du{G}_a(\vc{r},\vc{r}')\cdot\overline{\vc{F}^{(2)}_R}(\vc{r}')\;d\vc{r}'.
\end{equation}
By symmetry we can expand in vector spherical harmonics as
\begin{equation}
\label{5.5}\overline{\vc{F}^{(2)}_R}(\vc{r})=\varepsilon^2\rho \omega^2 a\bigg[\sum^\infty_{l=1}F_{RAl}(r)\vc{A}_l(\hat{\vc{r}})+
\sum^\infty_{l=0}F_{RBl}(r)\vc{B}_l(\hat{\vc{r}})\bigg],
\end{equation}
with dimensionless scalar amplitudes $F_{RAl}(r),\;F_{RBl}(r)$.
With the notation
\begin{equation}
\label{5.6}\langle\vc{V}_k|\vc{W}_l\rangle=\int^\pi_0\vc{V}_k\cdot\vc{W}_l\sin\theta\;d\theta
\end{equation}
the orthonormality relations of the harmonics are
\begin{equation}
\label{5.7}\langle\vc{A}_k|\vc{A}_l\rangle=2l\delta_{kl},\qquad
\langle\vc{A}_k|\vc{B}_l\rangle=0,\qquad
\langle\vc{B}_k|\vc{B}_l\rangle=(2l+2)\delta_{kl}.
\end{equation}
We note that $\vc{B}_0=-\vc{e}_r$.
In the present case we find that only amplitudes for $l=0$ and $l=2$ differ from zero. We read off
\begin{eqnarray}
\label{5.8}F_{RB0}(r)&=&\frac{-3s^3}{16u^7}\bigg[\bigg(3s+2s^2-e^{s-u}[(3u+2u^2)\cos(s-u)
+(3+3u)\sin(s-u)]\bigg)^2\nonumber\\
&+&\bigg(3+3s-e^{s-u}[(3+3u)\cos(s-u)-(3u+2u^2)\sin(s-u)]\bigg)^2\bigg],
\end{eqnarray}
where $u=sr/a$. In addition
\begin{eqnarray}
\label{5.8}F_{RA2}(r)=\frac{3s^3}{160u^7}&\bigg[&9+18s+18s^2+12s^3+4s^4+(9+18u+18u^2+12u^3+4u^4)e^{2s-2u}\nonumber\\
&-&(18+18s+18u+12u^3+36su+12s^2u-8s^2u^3)e^{s-u}\cos(s-u)\nonumber\\
&-&(18s+12s^2-18u+12u^3+12s^2u+24su^3+8s^2u^3)e^{s-u}\sin(s-u)\bigg].\nonumber\\
\end{eqnarray}
Finally
\begin{eqnarray}
\label{5.9}F_{RB2}(r)=\frac{-s^3}{20u^7}&\bigg[&9+18s+18s^2+12s^3+4s^4+(9+18u+18u^2-3u^3-6u^4)e^{2s-2u}\nonumber\\
&-&(18+18s+18u-3u^3+36su+12s^2u+2s^2u^3)e^{s-u}\cos(s-u)\nonumber\\
&-&(18s+12s^2-18u-3u^3+12s^2u-6su^3-2s^2u^3)e^{s-u}\sin(s-u)\bigg].\nonumber\\
\end{eqnarray}
The expression for $F_{RB0}(r)$ can be cast in a form similar to Eqs. (5.9) and (5.10), but the present form makes clear that $F_{RB0}(r)$ is negative definite. This contribution does not give rise to a flow, since the Green function for $l=0$ vanishes, as can be seen from Eq. (4.7). This contribution to the Reynolds force density is compensated by a pressure gradient, as we shall discuss in Sec. VIII.

The bulk contribution to the steady streaming velocity at any point $\vc{r}$ is now given by a sum of eight integrals labeled $ij2\sigma$, where $\sigma$ equals $<$ or $>$ and $i,j$ equals $A$ or $B$. The required radial integrals can all be evaluated, but the result is unwieldy. It is more straightforward to use numerical integration. The first contribution is
\begin{equation}
\label{5.11}\overline{\vc{u}^{(2)}}_B(\vc{r})_{(AA2<)}=\varepsilon^2 \frac{\rho \omega^2 a}{\eta}
\int_r^\infty G_{AA2<}(r,b)F_{RA2}(b)\;db\;\vc{A}_2(\hat{\vc{r}}),
\end{equation}
and the second is
\begin{equation}
\label{5.12}\overline{\vc{u}^{(2)}}_B(\vc{r})_{(AA2>)}=\varepsilon^2 \frac{\rho \omega^2 a}{\eta}
\int^r_aG_{AA2>}(r,b)F_{RA2}(b)\;db\;\vc{A}_2(\hat{\vc{r}}).
\end{equation}
The remaining terms are similar.

Our final result for the steady streaming velocity is
\begin{equation}
\label{5.13}\overline{\vc{u}^{(2)}}({\vc{r}})=\overline{\vc{u}^{(2)}}_S({\vc{r}})+\overline{\vc{u}^{(2)}}_B({\vc{r}}),
\end{equation}
with the first term given by Eq. (3.9) and the second by a sum of eight integrals like Eq. (5.11), which are conveniently calculated numerically. The velocity may be expressed as a sum of two vector spherical harmonics,
\begin{equation}
\label{5.14}\overline{\vc{u}^{(2)}}({\vc{r}})=\omega a\bigg[V_A(r)\vc{A}_2(\hat{{\vc{r}}})+V_B(r)\vc{B}_2\hat{({\vc{r}}})\bigg],
\end{equation}
with dimensionless radial functions $V_A(r)$ and $V_B(r)$.

In Figs. 1 and 2 we show the flow pattern in the first quadrant of the $xz$-plane for two values of the scale number, $s=1$ and $s=10$. The first value is smaller than the transition value $s_0=2.85632$, the second is larger. In both figures we have indicated the center of the vortex.

For $s<s_0$ the flow velocity on the positive $z$-axis is in the negative direction for all values of $z$. For $s>s_0$ it is in the positive direction for $z$ beyond the splitting point with $z$-coordinate $z_0(s)$, where $\overline{u_z^{(2)}}((0,0,z_0))=0$. In Fig. 3 we plot $z_0(s)$ as a function of $s$. For $s=10$ one has $z_0(10)=1.2855\;a$. The value diverges as $s$ tends to $s_0$ from above. Over the whole range shown in Fig. 3 the curve is well fitted by $2.6562/\sqrt{s-s_0}$. For small $s-s_0$ this changes to $2.584/(s-s_0)^\gamma$ with exponent $\gamma=0.425$.
 \\\\
We consider the value of the $z$-component of the steady streaming velocity $\overline{u_z^{(2)}}((0,0,z))$ on the positive $z$-axis in some more detail for several values of $s$. We choose $s={1,\;10,\;20,\;50}$, and compare with the result of Riley\cite{8} for large $s$. The curves in Fig. 4 are most easily visualized in terms of a potential picture. Each of the curves vanishes at $z=a$ and at infinity, and shows a potential well for small $s$. For very small $s$ the well is very shallow. As $s$ increases it deepens and narrows, with the minimum decreasing. This goes on until the number $s$ reaches the value $s_0=2.85632$. At this point a barrier appears at infinity, which grows in magnitude with a maximum which shifts to smaller values of $z$, as $s$ increases further. At $s=s_0$ the boundary layer extends to infinity. We calculate the critical value $s_0$ at the end of Sec. VI.

A similar plot can be considered for the $x$-component of the steady streaming velocity $\overline{u_x^{(2)}}((x,0,0))$ on the positive $x$-axis. On account of Eq. (5.14) one has the simple relation
\begin{equation}
\label{5.15}\overline{u_z^{(2)}}((0,0,z))=-2\overline{u_x^{(2)}}((z,0,0)),
\end{equation}
so that the plot follows immediately from Fig. 3. The flow velocity vanishes at the stagnation point $(x_0(s),0,0)$. On account of Eq. (5.15) one has $x_0(s)=z_0(s)$.

From the angular dependence in Eq. (5.14) we see that on the cone defined by the polar angle $\theta_0$, given by $P_2(\cos\theta_0)=0$ or
 \begin{equation}
\label{5.16}\theta_0=\frac{1}{2}\arccos\frac{-1}{3}=0.955317,
\end{equation}
the radial component of the velocity vector vanishes. The center of the vortex lies on this cone. At the center the $\theta$-component of the velocity vector vanishes also. It is interesting to note that the angle $54.75^\circ$ indicated in Fig. 3 of Wang's article \cite{7} is very close to $(\theta_0/\pi)\;180^{\circ}=54.7356^\circ$. The distance of the vortex center to the origin $r_0(s)$ is plotted in Fig. 5 as a function of $s$.
Particular values are $r_0(1)=1.7157\;a$ and $r_0(10)=1.0799\;a$. Over the whole range shown in Fig. 5 the curve is reasonably well fitted by $1/s^\gamma+0.696$ with exponent $\gamma=0.444$. For small $s$ this changes to $1.714/s^\gamma$ with exponent $\gamma=0.282$.

The transition point $s_0=2.85632$ clearly is special. In Fig. 6 we show the  streamlines for $s=3$. It appears that the boundary layer extends much farther than in Fig. 2 before the outer region is reached.

\section{\label{VI}Outer region}

In this section we study in particular the flow in the outer region, where the mean Reynolds force density can be neglected and the flow velocity and pressure satisfy the homogeneous Stokes equations. The outer flow is given by a sum of two basic flows, each weighted by a moment.

In the following we derive analytic expressions for the two moments $\kappa_{2B}$ and $\mu_{2B}$ which dominate the bulk flow at large distance. We note from the expressions (5.9) and (5.10) that the two Reynolds force densities take appreciable value in a region about the spherical surface $r=a$ of approximate width $\lambda=a/s$. In Fig. 7 we plot the two radial functions $F_{RA2}(r)$ and $F_{RB2}(r)$ for $s=3$ as functions of $u=sr/a$. If the functions were to have a cut-off distance $b_0$ beyond which they vanish, then integrals of the type (5.11) vanish identically for $r>b_0$, and integrals of the type (5.12) yield solutions of the steady-state Stokes equations. In the present case these are proportional to $\vc{v}^0_2(\vc{r})$ and $\vc{u}_2(\vc{r})$, given by Eq. (3.7).  We observe the two identities
\begin{eqnarray}
\label{6.1}
G_{AA2>}(r,b)\;\vc{A}_2(\hat{\vc{r}})+G_{BA2>}(r,b)\;\vc{B}_2(\hat{\vc{r}})&=&
a\big[C_{0A}\vc{u}_{20}(\vc{r})+C_{2A}\vc{u}_{22}(\vc{r})\big],\nonumber\\
G_{AB2>}(r,b)\;\vc{A}_2(\hat{\vc{r}})+G_{BB2>}(r,b)\;\vc{B}_2(\hat{\vc{r}})&=&
a\big[C_{0B}\vc{u}_{20}(\vc{r})+C_{2B}\vc{u}_{22}(\vc{r})\big],
\end{eqnarray}
with dimensionless vector functions
\begin{eqnarray}
\label{6.2}\vc{u}_{20}&=&\frac{a^2}{r^2}\;\big(\vc{A}_2-\vc{B}_2\big)=\frac{5}{3}\;\vc{v}^0_2,\nonumber\\
\qquad\vc{u}_{22}&=&\frac{a^4}{r^4}\;\vc{B}_2=-\vc{u}_2,
\end{eqnarray}
and dimensionless coefficients
\begin{eqnarray}
\label{6.3}C_{0A}&=&\frac{b^3-a^3}{5a^3},\qquad C_{2A}=\frac{b^5-a^5}{5a^5},\nonumber\\
C_{0B}&=&\frac{3(b^2-a^2)}{10b^2},\qquad C_{2B}=\frac{4b^7+21a^5b^2-25a^7}{70a^5b^2}.
\end{eqnarray}
The two identities imply a separation of the variables $b$ and $r$. The subscripts $2\sigma$ with $\sigma=0,2$ arise from the relation to the functions $\vc{v}^-_{l0\sigma}$ defined by Cichocki et al.\cite{23}.

It follows from Eqs. (5.12) and (6.1) that the far-field contribution to the bulk flow is given by
\begin{eqnarray}
\label{6.4}\overline{\vc{u}^{(2)}}_B(\vc{r})\approx\varepsilon^2\frac{\rho\omega^2 a^2}{\eta}\;\int^\infty_a &\big[&\big(C_{0A}(b)F_{RA2}(b)+C_{0B}(b)F_{RB2}(b)\big)\vc{u}_{20}(\vc{r})\nonumber\\
&+&\big(C_{2A}(b)F_{RA2}(b)+C_{2B}(b)F_{RB2}(b))\vc{u}_{22}(\vc{r})\big]\;db.\nonumber\\
\end{eqnarray}
By use of Eq. (6.2) this can be rewritten in the conventional form \cite{28}
\begin{equation}
\label{6.5}\overline{\vc{u}^{(2)}}_B(\vc{r})\approx -\omega a\big[\kappa_{2B}\vc{v}^0_{2}(\vc{r})
+\mu_{2B}\vc{u}_2(\vc{r})\big],
\end{equation}
similar to Eq. (3.9). Performing the integral over $b$ we find $\kappa_{2B}=\varepsilon^2 K_{2B}(s)$ with
\begin{eqnarray}
\label{6.6}K_{2B}(s)=\frac{-s}{480}&\bigg[&84+18s+4s^2-36s^3+10s^4-2s^5+2s^6\nonumber\\
&+&s^3\big[15i+(15+15i)s+13s^2+(3-3i)s^3-2is^4\big]F_+\nonumber\\
&+&s^3\big[-15i+(15-15i)s+13s^2+(3+3i)s^3+2is^4\big]F_--48s^5F_2\bigg],\nonumber\\
\end{eqnarray}
with the abbreviations
\begin{equation}
\label{6.7}F_+=F(s+is),\qquad F_-=F(s-is),\qquad F_2=F(2s),
\end{equation}
where the function $F(\zeta)$ with complex variable $\zeta$ is defined by
\begin{equation}
\label{6.8}F(\zeta)=e^\zeta E_1(\zeta)=\int^\infty_0\frac{e^{-u}}{\zeta+u}\;du.
\end{equation}
Similarly we find $\mu_{2B}=\varepsilon^2 M_{2B}(s)$ with
\begin{eqnarray}
\label{6.9}M_{2B}(s)=\frac{1}{3360s}&\bigg[&378+756s+660s^2+102s^3+12s^4-156s^5+58s^6-10s^7+10s^8\nonumber\\
&+&s^5\big[63i+(63+63i)s+57s^2+(15-15i)s^3-10is^4\big]F_+\nonumber\\
&+&s^5\big[-63i+(63-63i)s+57s^2+(15+15i)s^3+10is^4\big]F_--240s^2F_2\bigg].\nonumber\\
\end{eqnarray}
There is a pronounced dependence on the scale number $s=a\sqrt{\rho\omega/2\eta}$. We show below that the divergent behavior at large $s$ of the surface moments, shown in Eq. (3.10), is precisely canceled by the bulk contribution. In Fig. 8 we show the behavior of the functions
\begin{equation}
\label{6.10}K_2(s)=\frac{1}{4}s+K_{2B}(s),\qquad M_2(s)=\frac{-1}{4}s+M_{2B}(s)
\end{equation}
as functions of $\log_{10}s$.

The plots shown in Fig. 8 actually consist of three parts corresponding to different calculations of the moment functions $K_2(s)$ and $M_2(s)$. In the interval $0<s<0.3$ we calculate the functions $F_+,\;F_-$ and $F_2$, defined in Eq. (6.7), from a truncation of the continued fraction representation\cite{29} of the exponential integral function $E_1(\zeta)$. In the interval $0.3<s<25$ the functions are calculated by numerical integration, and for $s>25$ the functions are calculated from the asymptotic expansion of the exponential integral function.

We discuss the behavior of the moment functions $K_2(s)$ and $M_2(s)$ for small and large $s$ in some more detail. Expanding the function  $K_2(s)$ about $s=0$ we find for small $s$
\begin{equation}
\label{6.11}K_2(s)=\frac{3s}{40}-\frac{3s^2}{80}-\frac{s^3}{120}+O(s^4).
\end{equation}
Similarly we find for the function $M_2(s)$
\begin{equation}
\label{6.12}M_2(s)=\frac{9}{80s}+\frac{9}{40}-\frac{3s}{56}+\frac{17s^2}{560}+O(s^3).
\end{equation}
For large $s$ we find for the function $K_2(s)$ the expansion
\begin{equation}
\label{6.13}K_2(s)=-\frac{15}{16}+\frac{189}{16s}-\frac{2571}{32s^2}+\frac{2343}{8s^3}+O(s^{-4}),
\end{equation}
and for the function $M_2(s)$
\begin{equation}
\label{6.14}M_2(s)=\frac{15}{16}-\frac{69}{8s}+\frac{1881}{32s^2}-\frac{1701}{8s^3}+O(s^{-4}).
\end{equation}
Hence for both moments the divergence of the surface term for large $s$, shown in Eq. (3.10), is canceled precisely by a corresponding cancelation from the bulk term. We found a similar cancelation in the mean swimming velocity of a deforming sphere\cite{19,30}. The cancelation was found also for the mean swimming velocity of an oscillating two-sphere, as shown by Derr et al.\cite{31}.

The first term in the expansion (6.13) for the moment $K_2(s)$ and in the expansion (6.14) for the moment $M_2(s)$ is
equivalent to the result at high frequency found by Riley\cite{8} and verified by Spelman and Lauga\cite{14}. Riley's expression for the mean stream function reads in our present notation
\begin{equation}
\label{6.15}\overline{\psi}_R(r,\theta)=-\frac{45}{32}\varepsilon^2 \omega a\bigg(\frac{a^2}{r^2}-1\bigg)
\sin^2\theta\cos\theta.
\end{equation}
Hence one derives the corresponding mean flow velocity
\begin{equation}
\label{6.16}\overline{\vc{u}}(r,\theta)=\frac{15}{16}\;\varepsilon^2 \omega a\;\big[\vc{v}^0_2(\vc{r})-\vc{u}_2(\vc{r})\big].
\end{equation}
Our result extends this to all values of $s$ as
\begin{equation}
\label{6.17}\overline{\vc{u}}(r,\theta)\approx -\varepsilon^2 \omega a\;\big[K_2(s)\vc{v}^0_2(\vc{r})+M_2(s)\vc{u}_2(\vc{r})\big].
\end{equation}
In particular the mean velocity along the positive $z$-axis for large $z$ is given by
\begin{equation}
\label{6.18}\overline{u_z}(z)\approx -3\varepsilon^2 \omega a\;\big[K_2(s)\frac{a^2}{z^2}+M_2(s)\frac{a^4}{z^4}\big].
\end{equation}
 The expressions (6.17) and (6.18) are exact to second order in $\varepsilon$, but hold only for sufficiently large $r$. We find numerically that Eq. (6.17) provides a good approximation to the exact value given by Eq. (5.14) provided $r>2a$. The behavior shown in Figs. 1, 2, and 6 was calculated from Eq. (5.14).

 Riley\cite{8} derived an expression for the steady streaming velocity also in the low frequency limit. The expression involved a matched asymptotic expansion. The two figures presented by Riley\cite{8} for the low and high frequency limits show different flow directions in the two cases. This suggests the existence of a flow transition at some intermediate value of the frequency. We find the flow reversal at the single zero $s_0=2.85632$ of the function $K_2(s)$. For $s<s_0$ the mean flow velocity is negative for all points on the positive $z$-axis.

\section{\label{VII}Pumping}

 The steady flow pattern shows that the oscillating sphere may be regarded as a pump. For large $r$ the velocity vector points in the radial direction and for $0<s<s_0$ and $0<\theta<\theta_0$ is directed inward, whereas for $\theta_0<\theta<\pi/2$ it is directed outward. For $s_0<s<\infty$ these directions are reversed. Hence for $0<s<s_0$ the oscillating sphere can be regarded as a pump moving fluid from the polar areas $\theta\approx 0$ and $\theta\approx\pi$ into the equatorial plane $\theta\approx\pi/2$. For $s_0<s$ the pump operates in the opposite direction. The mass flow rate is given by
 \begin{equation}
\label{7.1}\overline{q}_m=4\pi\rho\lim_{r\rightarrow\infty}\int^{\theta_0}_0|\overline{u_r}|r^2\sin\theta\;d\theta.
\end{equation}
From Eq. (6.17) we find
 \begin{equation}
\label{7.2}\overline{q}_m=\varepsilon^2A_f\;|K_2(s)|\rho\omega a^3,
\end{equation}
with the angular factor
\begin{equation}
\label{7.3}A_f=6\pi\int^{\theta_0}_0P_2(\cos\theta)\sin\theta\;d\theta=3.62760.
\end{equation}
The volumetric flow rate is $\overline{q}_v=\overline{q}_m/\rho$. The flow rate vanishes at the critical value $s_0$. The dependence on $s$ is shown in Fig. 8.

We define the efficiency of pumping by comparing the rate of flow with the rate of dissipation. The second order mean rate of dissipation is given by \cite{20}
 \begin{equation}
\label{7.4}\overline{\mathcal{D}_2}=2\eta\int(\nabla\vc{u}^{(1)*}_\omega)^s
: (\nabla\vc{u}^{(1)}_\omega)^s\;d\vc{r},
\end{equation}
where the superscript $s$ indicates symmetrization of the tensor. Alternatively the mean rate of dissipation may be evaluated from a surface integral. We find
\begin{equation}
\label{7.5}\overline{\mathcal{D}_2}=6\pi(1+s)\varepsilon^2\eta\omega^2a^3.
\end{equation}
The corresponding efficiency is a dimensionless measure of the ratio of benefit and cost. In analogy to the theory of swimming\cite{16} we define it here as the ratio $L=8\pi\eta\omega a^2\overline{U}/\overline{\mathcal{D}_2}$, where $\overline{U}$ is a typical flow velocity. We define this from the mass flow rate by putting
\begin{equation}
\label{7.6}\overline{q_m}=4\theta_0a^2\rho\overline{U},
\end{equation}
where the prefactor $4\theta_0$ is chosen such that it would be $2\pi$ if $\theta_0$ were $\pi/2$.
Then the efficiency becomes
\begin{equation}
\label{7.7}L=\frac{A_f}{3\theta_0}\;\frac{|K_2(s)|}{1+s}.
\end{equation}
For $s<s_0$ the function $|K_2(s)|/(1+s)$ takes the maximum value $0.0202$ at $s_m=0.7770$, and for $s>s_0$ it takes the maximum value $0.0268$ at $s_p=12.188$.

\section{\label{VIII}Steady pressure}

Besides the steady streaming velocity we can also calculate the steady deviation from the ambient equilibrium pressure $p_0$. By axial symmetry the steady mean pressure $\overline{p}(r,\theta)$ can be expanded in a series of Legendre polynomials $P_l(\cos\theta)$. In the present case only the terms $l=0$ and $l=2$ contribute to second order in the amplitude . The isotropic part $\Delta p(r)$ is driven by the isotropic contribution to the Reynolds force density found in Eq. (5.8). To second order in the amplitude we find from Eq. (2.4)
\begin{equation}
\label{8.1}\nabla^2\overline{p^{(2)}}=\nabla\cdot\overline{\vc{F}_R^{(2)}}.
\end{equation}
The corresponding ordinary differential equation for the isotropic part is easily integrated.

The calculation of the anisotropic part, proportional to $P_2(\cos\theta)$, can be performed in parallel to that of Sec. IV. We recall from Eq. (2.7) that the steady pressure is the sum of a surface and a bulk contribution. The surface contribution corresponds to Eq. (3.9), and is given by
\begin{equation}
\label{8.2}\overline{p_S^{(2)}}=\frac{-1}{4}\;\varepsilon^2\;s\;\omega a\;p^0_2(\vc{r}),
\end{equation}
with the function
\begin{equation}
\label{8.3}p^0_2(\vc{r})=2\eta\;\frac{a^2}{r^3}\;P_2(\cos\theta).
\end{equation}
The bulk contribution is generated by the mean Reynolds force density acting in the fluid.

In Eq. (4.4) there appear four vector functions which represent flows generated by a delta-function force density located on a spherical shell of radius $b$ surrounding a sphere of radius $a$ on which the flow satisfies the no-slip boundary condition. The fluid is viscous and incompressible, and the flow velocity satisfies the Stokes equations of steady flow together with a scalar pressure function. The latter can be found by acting with the pressure part of the Green function on the mean Reynolds force density

We construct the scalar pressure functions corresponding to the eight flow velocities in Eq. (4.4), specializing to $l=2$ for simplicity. Corresponding to the list in Eq. (4.8) we find
\begin{eqnarray}
\label{8.4}Q_{AA2<}(r,b)&=&0,\qquad Q_{AA2>}(r,b)=0,\nonumber\\
Q_{AB2<}(r,b)&=&\frac{3r^2}{b^2},\qquad Q_{AB2>}(r,b)=0,\nonumber\\
Q_{BA2<}(r,b)&=&\frac{-2a^3}{r^3},\qquad Q_{BA2>}(r,b)=\frac{2(b^3-a^3)}{r^3},\nonumber\\
Q_{BB2<}(r,b)&=&\frac{3a^3(b^2-a^2)}{b^2r^3},\qquad Q_{BB2>}(r,b)=\frac{3a^3(b^2-a^2)}{b^2r^3}.
\end{eqnarray}
In analogy to Eqs. (5.11) and (5.12) the bulk contribution to the steady pressure follows from a sum of five integrals, the first of which is
\begin{equation}
\label{8.5}\overline{p^{(2)}}_B(\vc{r})_{(AB2<)}=\varepsilon^2 \rho \omega^2
\int_r^\infty Q_{AB2<}(r,b)F_{RB2}(b)\;db\;P_2(\cos\theta).
\end{equation}
The remaining terms are similar.

\section{\label{IX}Conclusion}

In the above we found the analytic expression for the steady streaming flow pattern about an oscillating sphere. To second order in the amplitude of oscillations the mean flow velocity takes the form Eq. (5.14) with complicated radial functions. The derived integral expressions for these functions can be readily evaluated numerically. Typical flow patterns are shown in Figs. 1, 2, and 6.

The presented derivation is based on a technique involving a new form of antenna theorem \cite{19}. Here we have used a form valid for no-slip boundary conditions on a sphere immersed in an incompressible fluid, but similar theorems can be derived for other boundary conditions, for compressible fluids, and other geometries. Thus we expect that the present calculation can serve as an example for a wide field of applications.

Mathematically our approach is rather different from that used traditionally for this problem. Traditionally one uses the axial symmetry to reduce the problem to a scalar one by the introduction of a stream function \cite{8,9,14}. On the contrary, by expanding in vector spherical harmonics we make sure that proper account is taken of the rotational symmetry \cite{24}. We find that only orders $l=0$ and $l=2$ are needed, whereas in the stream function approach one requires an infinite sum, and approximates by truncation.

Amin and Riley \cite{9} claim that they solve the present problem, but in fact their geometry is rather different. They consider a sphere of radius $a$ with center fixed at the origin, and an oscillatory point source at height $R$ on the $z$-axis generating the flow. They claim that the limit where both $R$ and the strength of the source tend to infinity corresponds to the present problem. The figures suggest that this is correct, but the limit procedure is not mathematically attractive.

The single sphere flow pattern, as derived here, has an interest of its own. It shows an interesting dependence on the scale parameter $s=a\sqrt{\omega\rho/2\eta}$, with a flow reversal at the critical value $s_0=2.85632$. Clearly this will have drastic effects in the two-body problem, as shown in the analysis of the two-sphere swimming problem  by Derr et al. \cite{31}. In particular we expect that the reversal of swim direction seen by Dombrowski et al. \cite{32} is related to the flow reversal of the single sphere steady streaming flow. The single sphere flow velocity will be an equally important element for the understanding of many-sphere situations, as studied by Voth et al. \cite{11}.\\\\

The author has no conflicts of interest.



\newpage
\setlength{\unitlength}{1cm}
\begin{figure}
 \includegraphics{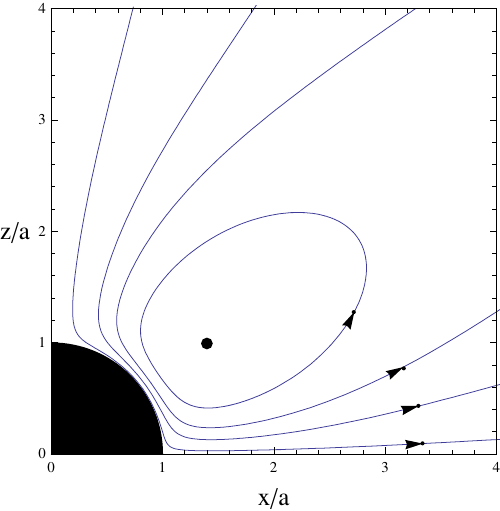}
   \put(-9.1,3.1){}
\put(-1.2,-.2){}
  \caption{\label{1}Plot of the steady streaming flow field $\overline{\vc{u}}(\vc{r})$ in the $xz$-plane for $s=1$.}
\end{figure}
\newpage
\clearpage
\newpage
\setlength{\unitlength}{1cm}
\begin{figure}
 \includegraphics{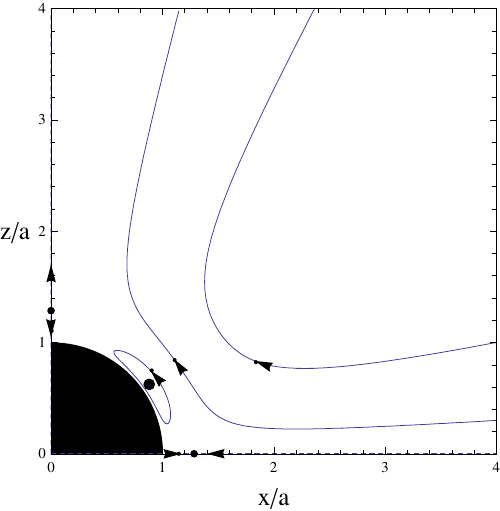}
   \put(-9.1,3.1){}
\put(-1.2,-.2){}
  \caption{\label{2}Plot of the steady streaming flow field $\overline{\vc{u}}(\vc{r})$ in the $xz$-plane for $s=10$.}
\end{figure}
\newpage
\clearpage
\newpage
\setlength{\unitlength}{1cm}
\begin{figure}
 \includegraphics{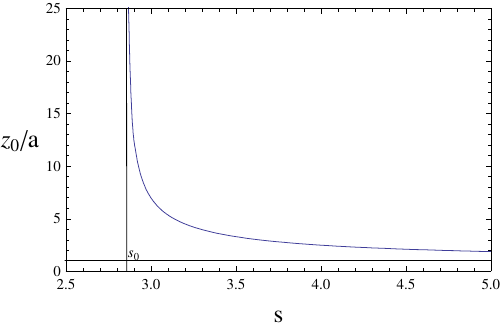}
   \put(-9.1,3.1){}
\put(-1.2,-.2){}
  \caption{\label{3} Plot of the splitting point $z_0(s)$ on the $z$-axis, and of the stagnation point $x_0(s)=z_0(s)$ on the $x$-axis, as functions of $s$.}
\end{figure}
\newpage
\clearpage
\newpage
\setlength{\unitlength}{1cm}
\begin{figure}
 \includegraphics{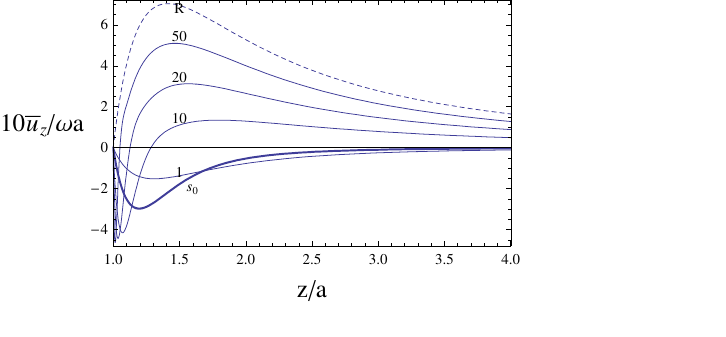}
   \put(-9.1,3.1){}
\put(-1.2,-.2){}
  \caption{\label{4}Plots of the velocity $\overline{u_z}(0,0,z)$ on the $z$-axis for various values of $s$.}
\end{figure}
\newpage
\clearpage
\newpage
\setlength{\unitlength}{1cm}
\begin{figure}
 \includegraphics{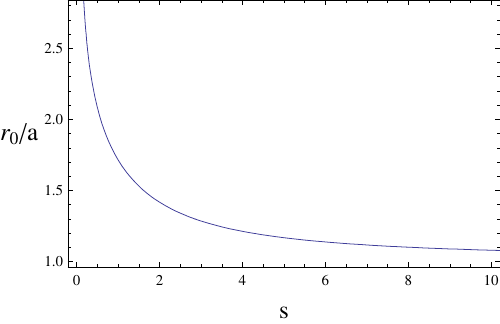}
   \put(-9.1,3.1){}
\put(-1.2,-.2){}
  \caption{\label{5}The vortex center is at distance $r_0(s)$ from the origin and at polar angle $\theta_0$ given by Eq. (5.16).}
\end{figure}
\newpage

\clearpage
\newpage
\setlength{\unitlength}{1cm}
\begin{figure}
 \includegraphics{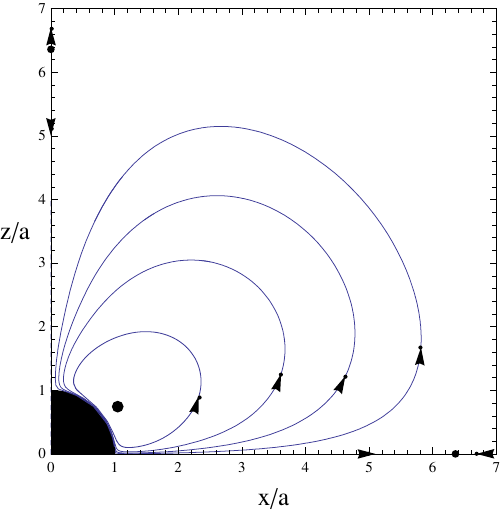}
   \put(-9.1,3.1){}
\put(-1.2,-.2){}
  \caption{\label{6}Plot of the steady streaming flow field $\overline{\vc{u}}(\vc{r})$ in the $xz$-plane for $s=3$.}
\end{figure}
\clearpage
\newpage
\setlength{\unitlength}{1cm}
\begin{figure}
 \includegraphics{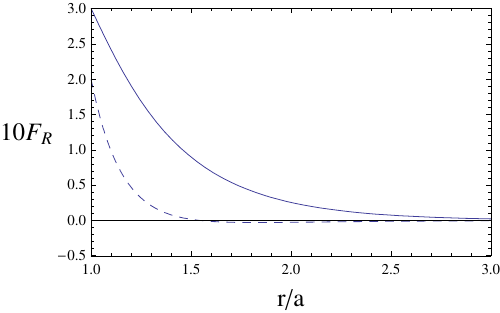}
   \put(-9.1,3.1){}
\put(-1.2,-.2){}
  \caption{\label{7}Plot of the radial functions $10F_{RA2}(r)$ (solid curve) and $10F_{RB2}(r)$ (dashed curve) for $s=3$ as functions of $r/a$.}
\end{figure}
\newpage
\newpage
\clearpage
\newpage
\setlength{\unitlength}{1cm}
\begin{figure}
 \includegraphics{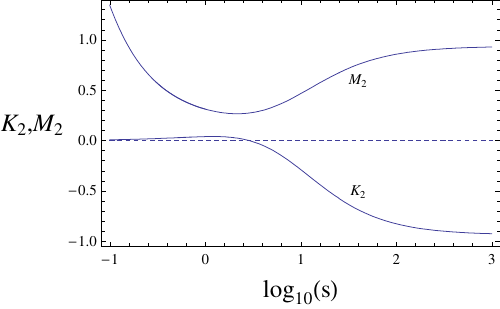}
   \put(-9.1,3.1){}
\put(-1.2,-.2){}
  \caption{\label{8}Plot of the functions $K_2(s)$ and $M_2(s)$.}
\end{figure}
\newpage

\end{document}